\documentclass[twocolumn,showpacs,preprintnumbers,amsmath,amssymb,aps]{revtex4-1}
\usepackage{graphicx}
\usepackage{dcolumn}
\usepackage{bm}

\begin{document}

\title{Concomitant enhancement of electron-phonon coupling and electron-electron interaction in graphene decorated with ytterbium}

\author{Minhee Kang$^{1}$}
\author{Jinwoong Hwang$^{1,2}$}
\author{Ji-Eun Lee$^{1,3}$}
\author{Alexei Fedorov$^{2}$}
\author{Choongyu Hwang$^{1}$}\email{ckhwang@pusan.ac.kr}

\affiliation{$^1$ Department of Physics, Pusan National University, Busan 46241, South Korea}
\affiliation{$^2$ Advanced Light Source, Lawrence Berkeley National Laboratory Berkeley, California 94720, USA}
\affiliation{$^3$ Center for Spintronics, Korea Institute of Science and Technology, Seoul 02792, South Korea}

\begin{abstract}
The interplay between electron-electron interaction and electron-phonon coupling has been one of the key issues in graphene as it can provide information on the origin of enhanced electron-phonon coupling in graphene by foreign atoms. In ytterbium-decorated graphene on SiC substrate, electron-phonon coupling exhibits strong enhancement compared to that of as-grown graphene. Based on angle-resolved photoemission study, the presence of ytterbium is also found to result in the decrease of Fermi velocity, revealing the enhancement of electron-electron interaction within the Fermi liquid theory. Our finding on the concomitant enhancement of electron-electron interaction and electron-phonon coupling suggests a possibility of the interplay between the two representative many-body interactions  in graphene decorated with foreign atoms.
\end{abstract}

\maketitle

\section{Introduction}

Decoration with foreign atoms has been one of the plausible methods to modify a solid-state surface and hence create/manipulate its properties. Even graphene, one of the inert surfaces self-assembled on the surface of SiC substrate, has been a playground to search for a variety of novel phenomena that do not exist when it stands alone as listed in Table~1. For example, the decoration leads to metal-insulator transition by NO$_2$~\cite{ShuyunPRL} and hydrogen~\cite{Aaron}, Kondo effect by cobalt~\cite{Ren} and cerium~\cite{Jinwoong}, magnetic effects by sulfur~\cite{Sulfur}, strongly enhanced electron-phonon coupling by alkali metals~\cite{Fedorov,Damascelli}. Especially, the electron-phonon coupling varies significantly depending on dopants with a dimensionless electron-phonon coupling constant ranging from 0.11 to 0.17 along the $\Gamma$K direction of the graphene unit cell and from 0.15 to 0.40 along the $\Gamma$M direction~\cite{Fedorov}. When lithium is adsorbed on graphene, the coupling constant reaches as high as 0.58 along a specific direction close to the $\Gamma$M direction that can possibly result in superconductivity~\cite{Damascelli}.

\begin{figure*}[]
\centering
\includegraphics[width=2\columnwidth]{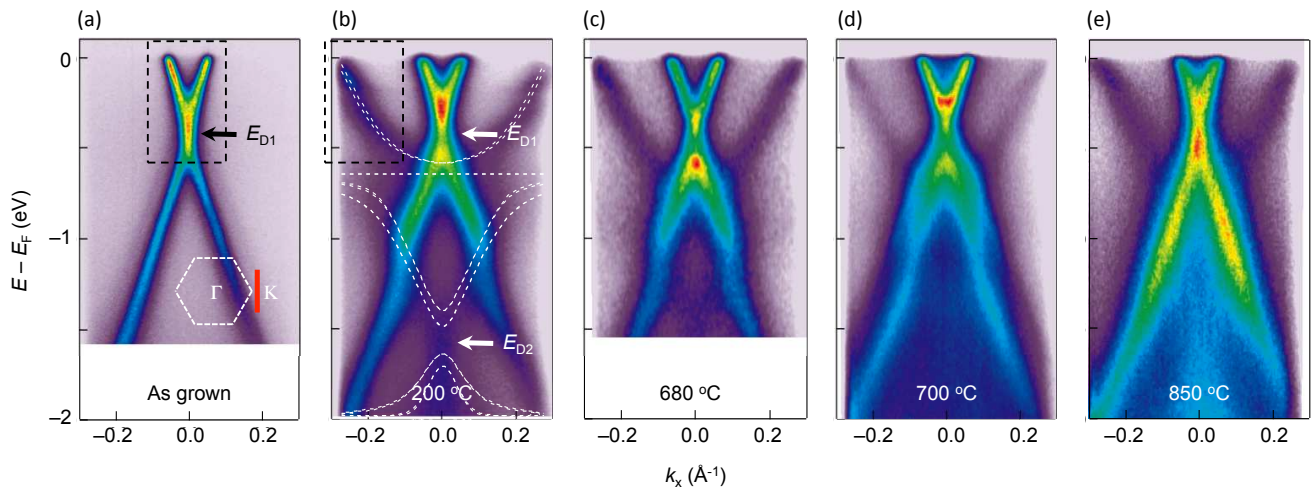}
\caption{\label{fig:fig1} (Color online) (a) An ARPES intensity map of as-grown graphene on an SiC(0001) substrate taken perpendicular to the $\Gamma$K direction of the graphene unit cell denoted by a red line and a white dashed hexagon, respectively, using 50 eV photons. (b-e) ARPES intensity maps of ytterbium-decorated graphene on an SiC(0001) substrate annealed at several different temperatures ranging from 200~$^{\circ}$C to 850~$^{\circ}$C taken perpendicular to the $\Gamma$K direction of the graphene unit cell denoted in the inset of panel (a). The white dashed curves in panel (b) are calculated band structure within the density functional theory for YbC$_6$, taken from Ref.~\cite{HwangYb}.}
\end{figure*}

Typically the strength of electron-phonon coupling changes as a function of charge carrier density. 
Undoped graphene has Fermi energy, $E_{\rm F}$, that is aligned with the crossing point between conduction and valence bands of graphene, so-called Dirac energy, $E_{\rm D}$. In this case, the number of electrons that can excite phonons are nominally zero, so that electron-phonon coupling is significantly suppressed. With increasing charge carrier density, the electron-phonon coupling in graphene increases with increasing charge carrier density, i.\,e.\,, in electron-doped graphene~\cite{Calandra_resolution}. 
However, calculated strength of electron-phonon coupling differs from the measured one in electron-doped graphene on substrates by as much as a factor of 5~\cite{Calandra_resolution}. While different substrates can result in different electronic/chemical environments beyond the simple change of charge carrier density, the discrepancy has been attributed to experimental artifacts such as a finite-resolution effect~\cite{Calandra_resolution} and a curved band structure which gives apparent enhancement especially along the $\Gamma$M direction~\cite{Calandra_BandStructureEffect,Park_vanHove}. On the other hand, experimentally determined strength is in excellent agreement with the theoretical values when graphene is placed on a metallic substrate such as Cu~\cite{DavidNJP}. This result gives a clue to understand the discrepancy, because a metallic substrate can efficiently screen electron-electron interaction in overlying graphene

\begin{table}[b]
\begin{center}
    \begin{tabular}{ | c | c | c | c | c | c | }
    \hline
    Phenomena  & MIT & Kondo & Magnetism & EPC & SCs\\ \hline 
    Decorations & NO$_2$~\cite{ShuyunPRL} & Co~\cite{Ren} & S~\cite{Sulfur} & AMs~\cite{Fedorov} & Li~\cite{Damascelli} \\
     & H~\cite{Aaron} & Ce~\cite{Jinwoong} & & Yb~\cite{HwangYb} & \\ \hline
    \end{tabular}
\end{center}
\caption{A variety of phenomena induced by the decoration of graphene with foreign atoms/molecules. MIT: metal-insulator transition, EPC: electron-phonon coupling, SCs: superconductivities.}
\end{table}


The role of foreign atoms on the electron-electron interaction, however, is not straightforward. It is because foreign atoms not only change the charge carrier density of graphene by charge transfer that simply increases participants in the electronic correlations, but also modify the orbital character of charge carriers in graphene by hybridization with orbitals of their own electrons. Partial intercalation provides a solution to overcome this obstacle. When epitaxial graphene on SiC substrate is mildly heated with foreign atoms on it, they are intercalated underneath the graphene layer to form atomic islands leaving the remaining area as pristine graphene~\cite{Gao}. The inhomogeneous intercalation results in the electron band structure of both heavily electron-doped graphene by foreign atoms and lightly electron-doped graphene by the SiC substrate, that are electronically coupled each other~\cite{HwangYb}. Although the lightly electron-doped graphene is influenced by the intercalants, it exhibits similar charge carrier density as pristine graphene, making it possible to investigate the change of electron-electron interaction modified by the presence of foreign atoms in the system.

Here we report the first experimental evidence of enhanced electron-electron interaction in graphene decorated by foreign atoms, in which strongly enhanced electron-phonon coupling is observed using angle-resolved photoemission spectroscopy. In the presence of ytterbium, graphene shows the electron band structure of both heavily and lightly electron-doped graphene, whose electronic coupling is modified by annealing at different temperatures. The former exhibits a peak-dip-hump structure due to enhanced electron-phonon coupling, whereas such spectral feature is not clearly observed in as-grown graphene, consistent with previous result~\cite{HwangYb}. The latter shows decreased Fermi velocity, $v_{\rm F}$, compared to that of as-grown graphene, revealing that electron-electron interaction is also enhanced by the presence of ytterbium within the Fermi liquid theory~\cite{Mahan}. As a result, our findings suggest that indeed electron-electron interaction can be responsible for the enhanced electron-phonon coupling observed in metal-adsorbed graphene.

\section{ARPES measurements}

Graphene samples were prepared by the epitaxial growth method on SiC(0001) substrate via silicon sublimation by e-beam heating~\cite{Rolling}. Ytterbium atoms were deposited on graphene at 100~K, followed by repeated annealing process at higher temperatures ranging from 200~$^{\circ}$C to 850~$^{\circ}$C. ARPES measurements were performed at 15~K using 50~eV photons at beamline 12.0.1~\cite{BL12} of the Advanced Light Source in Lawrence Berkeley National Laboratory. Energy and momentum resolutions were 32~meV and 0.01~\AA$^{-1}$. Sample orientation is  determined directly by finding the Brillouin zone corner, K-point, using the ARPES measurement.

\begin{figure*}[t]
\centering
\includegraphics[width=2\columnwidth]{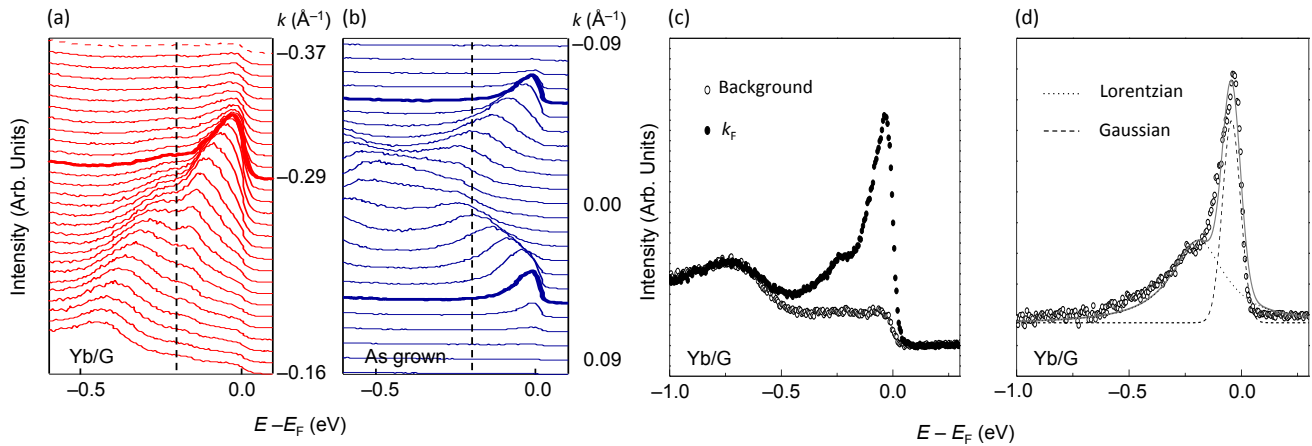}
\caption{\label{fig:fig2} (Color online) (a-b) Energy distribution curves (EDCs) of (a) heavily electron-doped graphene by ytterbium and (b) as-grown graphene on an SiC(0001) substrate. Bold curves correspond to the EDCs at $k_{\rm F}$ of each case. Black dashed lines denote the dip position in the EDCs for heavily electron-doped graphene. Red dashed curve in panel (a) is taken as a background signal that is free from the electron band structure of heavily electron-doped graphene by ytterbium. (c) Filled and empty circles are EDCs denoted by bold and dashed curves in panel (a), respectively, that are normalized with respect to the spectral intensity at lower energies. (d) The background signal is subtracted from the EDC taken at $k_{\rm F}$ to show the characteristic peak-dip-hump structure. The dashed and dotted curves are Gaussian and Lorentzian fits to data, respectively. The solid line is a total fit to the data.}
\end{figure*}

\section{Results and discussions}
\subsection{ARPES intensity maps of graphene decorated with Yb}

Figure~1 shows ARPES intensity maps for as-grown graphene and ytterbium-decorated graphene prepared with annealing process at several different temperatures. These maps are measured using 50~eV photons perpendicular to the $\Gamma$K direction of the graphene unit cell, denoted by a red line and a white dashed hexagon in the inset of Fig.~1(a), respectively. As-grown graphene shows a single Dirac cone that is lightly electron-doped by the presence of the SiC(0001) substrate~\cite{Seyller}, so that $E_{\rm D}$ is observed at $\sim$0.4~eV below $E_{\rm F}$, as shown in Fig.~1(a). Upon introducing ytterbium, the graphene sample exhibits two Dirac cones as shown in Fig.~1(b), that have with relatively weak and strong spectral intensity. The comparison between the electron band structure with the weak spectral intensity and the calculated band structure of graphene with a close-packed atomic layer of ytterbium (the white dashed curves calculated within the density functional theory for YbC$_6$~\cite{HwangYb}) identifies the weak spectral intensity as the Dirac cone of heavily electron-doped graphene by ytterbium whose $E_{\rm D}$ is at $\sim$1.6~eV below $E_{\rm F}$. The Dirac cone close to $E_{\rm F}$ shows non-linear dispersion due to the hybridization with the 4$f$ state of ytterbium that appears as the white dashed line at 0.6~eV below $E_{\rm F}$ in the calculated bands.

Another Dirac cone with strong spectral intensity resembles that of lightly electron-doped as-grown graphene, especially with similar $E_{\rm D}$. Interestingly, it does not show a signature of strong hybridization with the ytterbium 4$f$ state observed in the Dirac cone of heavily electron-doped graphene, but shows a kinked structure at 1.1~eV below $E_{\rm F}$, i.\,e.\,, at the crossing points with the Dirac cone of heavily electron-doped graphene. Such a spectral feature denotes that lightly electron-doped graphene does not directly interact with ytterbium, but is electronically coupled to heavily electron-doped graphene with ytterbium. With increasing annealing temperature, the curved band structure close to $E_{\rm F}$ is gradually flattened as shown in Figs.~1(c) and 1(d), suggesting that the hybridization between the ytterbium 4$f$ state and the Dirac cone of heavily electron-doped graphene becomes weaker, possibly because the close-packed atomic islands of ytterbium start to segregate due to thermal activation by the annealing process. When the annealing temperature is as high as 850~$^{\circ}$C, the kink-like structure observed at $\sim$1.1~eV below $E_{\rm F}$ is strongly suppressed and the characteristic linearity of the Dirac cone is gradually recovered as shown in Fig.~1(e).

\subsection{Enhanced electron-phonon coupling by the presence of Yb}

The effect of ytterbium on the electron-phonon coupling can be examined by the comparison of the Dirac cone of graphene that is in direct contact to ytterbium, i.\,e.\,, heavily electron doped graphene, with that of as-grown graphene. The Dirac cone of heavily electron-doped graphene by ytterbium was reported to show the characteristics of electron-phonon coupling whose dimensionless coupling constant is as high as 0.43 in experiments and 0.51 in calculations, suggesting a possibility of photon-mediate superconductivity with a phase transition temperature of as high as 2.17~K~\cite{HwangYb}. Indeed, energy distribution curves for the ARPES intensity map of heavily electron-doped graphene taken from a black-dashed rectangle in Fig.~1(b) shows a peak-dip-hump structure as shown in Fig.~2(a) with a peak close to $E_{\rm F}$, a dip at $\sim$0.2~eV below $E_{\rm F}$ as denoted by a black-dashed line, and a hump at lower energy. Since graphene phonon dispersions have two Kohn anomalies~\cite{Piscanec}, electronic coupling to the phonon modes in graphene is observed at $\sim$0.20~eV~\cite{AaronNatPhys} or $\sim$0.15~eV~\cite{Shuyun_phonon} below $E_{\rm F}$. Indeed, this is the characteristic feature of electron-phonon coupling observed in the electron density of states~\cite{Hengsberger,Ding}. On the other hand, as-grown graphene does not show a clear signature of such spectral feature at the phonon energy as shown in Fig.~2(b), indicating weaker electron-phonon coupling~\cite{Shuyun_phonon}. Figure~2(c) shows the energy spectra taken at Fermi wavenumber, $k_{\rm F}$, and away from $k_{\rm F}$ as denoted by the thick and dashed curves in Fig.~2(a), respectively. The latter is subtracted from the former to remove a background signal that is shown in Fig.~2(d). The resultant line shape is well fitted by a Gaussian peak function for the sharp peak near $E_{\rm F}$ and a Lorentzian peak function for the broad peak at higher binding energy. Each peak corresponds to a coherent quasiparticle state at lower binding energy than the phonon energy and an incoherent quasiparticle state due to the scattering with the phonons, respectively~\cite{Ding}. Although a quasiparticle peak is typically expected to be fitted with a Lorentzian, a Gaussian best fits the actual line shape at low temperatures, suggesting, near $E_{\rm F}$, random distribution of sharper peaks constitutes the qausiparticle peak~\cite{Ding}.

\subsection{Enhanced electron-electron interaction by the presence of Yb}

In the presence of enhanced electron-phonon coupling, the electron band structure of lightly electron-doped graphene provides information on the change of electron-electron interaction induced by the presence of ytterbium in the system. When charge carrier density changes, e.\,g.\,, electron-doped graphene versus undoped graphene, electronic correlations become distinctly different so that the former is approximated as a Fermi liquid system whereas the latter shows a clear signature of non-Fermionic behavior with strong electronic correlations~\cite{Sarma,HwangSR}. As a result, almost the same $E_{\rm D}$ of as-grown graphene and lightly electron-doped graphene in a ytterbium-decorated graphene sample makes it possible to compare the effect of the presence of ytterbium on electron-electron interaction within the Fermi liquid theory~\cite{Landau}. 

Figures~3(a) and~3(b) show ARPES intensity maps of as-grown graphene and a Lorentzian fit to its momentum distribution curves, respectively, resulting in the navy curves in Fig.~3(b)~\cite{HwangYb}. The slope of the energy-momentum dispersion for as-grown graphene is $v_{\rm F}=0.88\times10^6~{\rm m/s}$. The black curve is a tight-binding band that is fitted to the electron band structure of as-grown graphene, of which $E_{\rm D}$ is shifted to 0.4~eV below $E_{\rm F}$, consistent with previous calculations~\cite{Kim_gap}. Figure~3(c) shows an ARPES intensity map taken from lightly electron-doped graphene in the ytterbium-decorated graphene sample. Here the black curve is the same tight-binding band shown in Figs.~3(a) and~3(b) for comparison. Interestingly, the observed ARPES intensity gradually deviates from the tight-binding band upon approaching $E_{\rm F}$. To quantitatively compare the change of the electron band structure, Fig.~3(d) shows the result of a Lorentzian fit to the data. For ytterbium-decorated graphene, the slope of the energy-momentum dispersion is clearly different from the tight-binding band and hence the dispersion of as-grown graphene, providing $v_{\rm F}=0.75\times10^6~{\rm m/s}$ that is decreased by 15\% compared to that of as-grown graphene.

\begin{figure}[t]
\includegraphics[width=1\columnwidth]{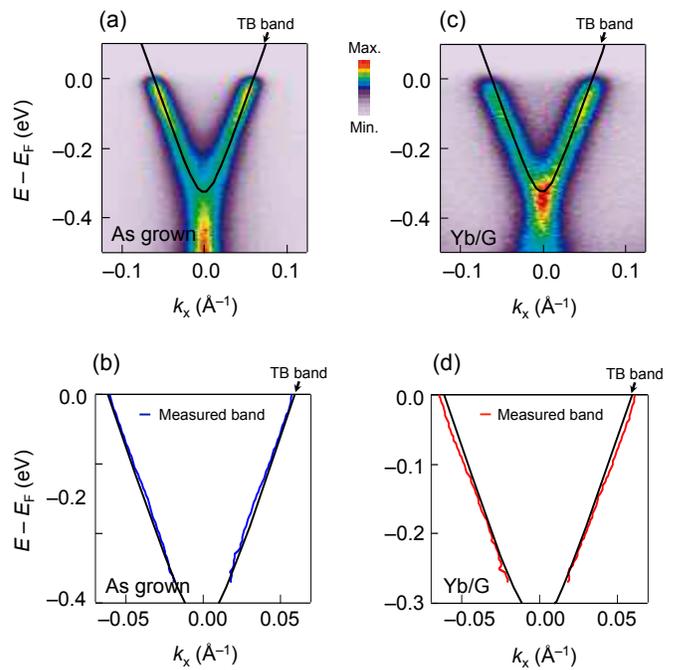}
\caption{\label{fig:fig3} (Color online) (a-b) An ARPES intensity map of as-grown graphene on an SiC(0001) substrate and a Lorentzian fit to the data to extract energy-momentum dispersion shown by navy curves~\cite{HwangYb}. (c-d) An ARPES intensity map of lightly electron-doped graphene in a ytterbium-decorated graphene sample and a Lorentzian fit to the data to extract energy-momentum dispersion shown by dark yellow curves. The black curve is a tight-binding band (TB band) that is fitted to the electron band structure of as-grown graphene. }
\end{figure}

The change of $v_{\rm F}$ provides a direct information on the change of electron-electron interaction. For quasiparticles with regular quadratic spectrum, e.\,g.\,, $E\,=\hbar^2 k^2/2m^*$, where $\hbar$ is Planck's constant, $k$ is wavenumber, and $m^*$ is effective mass of the quasiparticles, electron-electron interaction leads to the renormalization of their effective mass~\cite{Mahan}. In other words, as electron-electron interaction increases, effective mass becomes larger in Fermi liquids, so that the slope of the energy-momentum dispersion or the band width becomes narrower~\cite{Mi}. In undoped graphene, however, where the Dirac particles are regarded as massless particles with a linear spectrum, e.\,g.\,, $E\,=\,\hbar v_{\rm F}k$, velocity instead of mass is renormalized by electron-electron interaction~\cite{Sarma}. More specifically, as electron-electron interaction increases, $v_{\rm F}$ increases in contrast to what is expected in Fermi liquids, showing non-Fermi liquid behavior~\cite{HwangSR}. On the other hand, particles near $E_{\rm F}$ in electron-doped graphene is approximated as Fermi liquids~\cite{Sarma}. As a result, the decreased slope of the lightly electron-doped Dirac cone shown in Fig.~3(d) indicates that electron-electron interaction is enhanced by the presence of ytterbium in graphene. 

It is important to note that ytterbium is also the key factor in the strong enhancement of electron-phonon coupling in graphene~\cite{HwangYb}, although the exact role of ytterbium is not clear yet. A previous theoretical study predicts that with increasing Coulomb interaction beyond Hatree-Fock approximation, electronic coupling to the K point phonon ($\sim$0.16~eV) is enhanced whereas the coupling to the $\Gamma$ point phonon ($\sim$0.20~eV) stays the same~\cite{Basko}, which is similar to the observation of a peak in the real part of electron self-energy at $\sim$0.16~eV for the ytterbium-decorated graphene~\cite{HwangYb}. These previous theoretical~\cite{Basko} and experimental~\cite{HwangYb} works suggest that electron-electron interaction is enhanced by the presence of ytterbium, which is in turn responsible for the strong enhancement of electron-phonon coupling~\cite{HwangYb}. Indeed, our results shown in Fig.~3 provide an experimental evidence of the enhanced electron-electron interaction in graphene by the presence of ytterbium.

\section{Summary}

Ytterbium-decorated graphene exhibits the electron band structure of both heavily and lightly electron-doped graphene, corresponding to graphene with and without ytterbium in the same system, respectively, that are electronically coupled to each other. While the heavily electron-doped graphene shows the characteristics of enhanced electron-phonon coupling compared to as-grown graphene, lightly electron-doped graphene shows decreased $v_{\rm F}$ compared to that of as-grown graphene. This result provides not only an experimental evidence of enhanced electron-electron interaction within the Fermi liquid theory, but also important information to understand a possible role of electron-electron interaction on the enhancement of electron-phonon coupling in graphene by the presence of foreign atoms.

\section{Acknowledgements}
This work was supported by the National Research Foundation of Korea (NRF) grant funded by the Korea government (MSIT) (No.~2017K1A3A7A09016384 and No.~2018R1A2B6004538). JH acknowledges support from NRF-2017-Fostering Core Leaders of the Future Basic Science Program/Global Ph.D. Fellowship Program. The Advanced Light Source is supported by the Office of Basic Energy Sciences of the U.S. Department of Energy under Contract No. DE-AC02-05CH11231.

\end{document}